\documentstyle[aps,prl]{revtex}

\begin{document}
\draft
\title{Instabilities in Josephson Ladders with Current Induced Magnetic Fields}
\author{B. R. Trees and R. A. Murgescu}
\address{Department of Physics and Astronomy\\
Ohio Wesleyan University\\
Delaware, OH 43015}
\date{\today}
\maketitle
\begin{abstract}
We report on a theoretical analysis, consisting of both numerical and analytic work,
of the stability of synchronization of a ladder array of Josephson junctions under the
influence of current induced magnetic fields.  Surprisingly, we find that as the
ratio of the mutual to self inductance of the cells of the array is increased a region
of unstable behavior occurs followed by reentrant stable synchronization.  Analytic
work tells us that in order to understand fully the cause of the observed instabilities
the behavior of the vertical junctions, sometimes ignored in analytic analyses of ladder
arrays, must be taken into account.
\end{abstract}

\vspace{1em}
Ladder arrays of Josephson junctions are intriguing systems for a wealth of reasons: the possibility of phase locking, or synchronizing, a maximal subset of junctions suggests their use as microwave 
sources\cite{Barbara99}; they offer rich dynamical behavior, accessible to both theorists and experimentalists, in the field of coupled nonlinear 
oscillators (with recent interest in the prediction and observation of discrete 
rotobreathers)\cite{Binder00Trias00};   their complexity is between that of better understood one dimensional serial and parallel arrays and full 2D arrays ({\em e. g.} 
square arrays) and thus they offer a nice link between the two geometries; and ladder arrays 
can, under circumstances that are partially understood, be modeled by the discrete sine-Gordon (DSG) equation\cite{Ryu96}, which is itself a source of research 
interest among many\cite{Mingaleev00}.  With a desire to understand better the conditions
under which stable synchronization can occur,
we consider ladder arrays biased with uniform dc bias currents $I_B$ greater than 
the critical currents of the junctions and include the effects of current induced
magnetic fields (CIMF's) via self and mutual inductances of the cells of the array.  
(see Fig.~\ref{ladder})
Since the array is current-biased
above the critical current there will be a nonzero voltage across some subset
of junctions in the array.  These ``active'' junctions are synchronized, or phase locked, if
their voltages, after some initial transients, are identical functions of time. Furthermore,
previous workers have established that the effects of CIMF's may be important\cite{Majhofer91}
in determining the static and dynamics properties of arrays, and so it is natural to consider the effects
of CIMF's on synchronization as well.

\vspace{1em}
In this Letter we present numerical and analytical evidence that mutual inductance between
cells of a ladder array can lead to rich dynamical behavior, including destabilization of
synchronization and reentrant synchronization as the relative size of the mutual to self
inductance of the cells is increased.
The instability results for a finite range of values of the mutual inductance
and occurs when, for
the resistively and capacitively shunted junction (RCSJ) model,
the numerical solutions to the model equations diverge with time.  In addition to our numerical solutions we 
also investigate the dynamics analytically.  Namely, the two coupled RCSJ equations for the horizontal
and vertical junctions can be approximately decoupled.
The simplified equations allow us to calculate analytically the Floquet exponents, which measure the degree of stability of the synchronization.
Then, by direct comparison of the analytic results for the Floquet exponents, based on the simplified
equations, and the numerical results for the Floquet exponents, based on the full RCSJ equations, we learn valuable
information about the roles of the horizontal and vertical junctions in the array.  This technique
appears to be a powerful way to analyze the relative importance of subsets of junctions that experience different local conditions. 

\vspace{1em}
Josephson junctions consist of superconducting islands separated by a thin layer of 
nonsuperconducting material.  In the superconductors, the coherent motion of the paired
electrons, or Cooper pairs, leads to a wavefunction of the form
$\Psi =|\Psi|e^{i\theta}$, where $\theta$ is the macroscopic phase of the superconductor.  The
equations describing the dynamics of a single junction depend on the gauge-invariant
phase {\em difference}, or Josephson phase, across the junction, $\phi = \theta_1 - \theta_2 -(2\pi/\Phi_0)\int_{1}^{2}{\bf A}\cdot d{\bf l}$, where ${\bf A}$, the vector potential
due to an external magnetic field,  is integrated along a
path from one side of the junction to the other.  $\Phi_0 =\hbar/2e$ is the magnetic flux quantum, where
$\hbar$ is Planck's constant divided by $2\pi$ and $e$ is the electronic charge. We assume in this work that ${\bf A}=0$.


\vspace{1em}
Consider a ladder array of underdamped junctions with $N$ cells and periodic boundary
conditions, as shown in Fig.~\ref{ladder}.  
Each junction has a McCumber parameter $\beta_c\equiv 2\pi CI_{cx}R^2/\Phi_0$, where $C$($R$) is
the junction's capacitance (resistance), and  $I_{cx}$ is the critical current of a ``horizontal''
junction.
We assume each cell of four junctions is 
described by a self-inductance $L$ and also has a mutual inductance $-M$ ($M > 0$) with each of its two nearest neighbors, where $M < L$. As an adjustable parameter in the theory, $M/L$ is allowed to range between
zero and one.  Basic physics arguments for a simple ladder with
nearest neighbor mutual inductance precludes $M/L$ from exceeding $0.5$\cite{Phillips93}.
Nevertheless, it is informative and worthwhile to study the behavior of the model equations
over the range $0 < M/L < 1$.  Furthermore, such theoretical work could pique the interest
of others to try, using modern fabrication techniques, to enhance 
the mutual over the self inductance of the cells and thereby probe the broader range of $M/L$ values studied here.
Alternatively, one could point to these results for $M/L > 0.5$ as an interesting portent to
the behavior that might be observed in arrays with overlapping cells or in three dimensional arrays.

\vspace{1em}
An analysis based on the 
RCSJ equations and including the effect of induced magnetic flux leads to a pair of coupled equations
for the Josephson phases of the horizontal and vertical junctions,
\begin{equation}
\beta_c\phi^{\prime\prime} +\phi^{\prime} +\sin\phi + \frac{1}{\beta_L}\underline{Z}^{\mbox{Tr}}
\cdot\underline{X}^{-1}\cdot\left(\underline{Z}\cdot\phi + 2\psi\right)=0,
\label{eq:horizphases}
\end{equation}
\begin{equation}
\beta_c\psi^{\prime\prime} + \psi^{\prime} +\alpha\sin\psi +\frac{1}{\beta_L}\underline{X}^{-1}
\cdot\left(\underline{Z}\cdot\phi + 2\psi\right)=0,
\label{eq:vertphases}
\end{equation}
where $\beta_L\equiv 2\pi L I_{cx}/\Phi_0$ is the dimensionless self-inductance of a cell, and
$\alpha\equiv I_{cy}/I_{cx}$ is the critical current anisotropy.
The prime symbols denote differentiation with respect to dimensionless time, $\tau\equiv
t/t_c$ where $t_c\equiv hbar/2eI_{cx}R$.
These equations are compactly represented in matrix notation, where 
$\phi$ and $\psi$ are $N$-dimensional vectors representing the 
the Josephson phases of
the horizontal and vertical junctions, respectively.  $\underline{Z}$ is an
$N\times N$ matrix that depends on geometry, and $\underline{X}$ is the
dimensionless inductance matrix, also $N\times N$ in size.  The diagonal terms 
of $\underline{X}$ represent
the self-inductance of a given cell, {\em i.e.} $X_{jj}=+1$, while the mutual
inductance of the nearest neighbor cells is represented by the terms $X_{j,j\pm 1} =-\mu_L$,
where $\mu_L\equiv M/L$ is the dimensionless mutual inductance. All other elements of $\underline{X}$ are zero.
Eqs.~\ref{eq:horizphases} and~\ref{eq:vertphases} were solved numerically for $\phi$, $\psi$, 
$\phi^{\prime}$, and $\psi^{\prime}$ via a fourth-order Runge-Kutta algorithm as a function of the parameters $N$,
$\beta_c$, $\beta_L$, $\mu_L$, and $i_B\equiv I_B/I_{cx}$ (the dimensionless bias current).  
The starting configuration consisted of random
voltages and zero Josephson phases.

\vspace{1em}
As described elsewhere for a similar system\cite{Trees00} a stability analysis of the solutions to 
Eqs.~\ref{eq:horizphases} and~\ref{eq:vertphases} follows by letting $\phi=\phi_0+\eta$ and
$\psi=\psi_0+\delta$, where $\phi_0$ and $\psi_0$ are solutions to these equations.  
Eqs.~\ref{eq:horizphases} and~\ref{eq:vertphases} are linearized with respect to $\eta$
and $\delta$.   The time dependence of the
perturbations has the form $\eta\sim e^{\lambda t_c\tau}$ and $\delta\sim e^{\Lambda t_c\tau}$,
where $\lambda$ ($\Lambda$) represent the Floquet exponents for the horizontal (vertical) junctions.
If $\mbox{Re}(\lambda t_c) >0$ or $\mbox{Re}(\Lambda t_c) >0$ we expect unstable
behavior of the array for the given set of circuit parameters.  In fact, we are interested in
that exponent whose magnitude is closest to zero, as that describes the stability of the
longest-lived mode of the array.
 
\vspace{1em}
Figure~\ref{mutual}(a) shows the minimum Floquet exponent as a function of $\mu_L$ for
a 5-cell ladder.  
The three different sets of symbols represent three different sets of initial values for
the voltages, $\phi^{\prime}$ and $\psi^{\prime}$.
The data show that the
exponents do depend, to some degree, on the values of the starting voltages, although all the
runs follow the same trend. As $\mu_L$ is increased
from zero towards $0.5$, the stability of phase-locking increases, 
as shown by the negative exponent of growing magnitude, while the degree of stability decreases for
increasing $\mu_L$ greater than approximately $0.6$.  Even more interesting is the behavior
of the ladder in the range $0.5\leq\mu_L\alt 0.6$.  For these values of the mutual inductance
the ladder is actually unstable! This is evidenced by very rapidly growing phases and voltages
with time as Eqs.~\ref{eq:horizphases} and ~\ref{eq:vertphases} are numerically
integrated. For $N = 5$, the lower limit of this instability region is $\mu_{L}^{(1)}=0.5$
independent of other circuit parameters such as $\beta_c$ and $\beta_L$.  The upper limit of
this region, which we denote by $\mu_{L}^{(2)}$, depends on such quantities as the value of
the starting voltages as well as on the value of $\beta_L$.  For example, for a fixed set of
starting voltages, we find that $\mu_{L}^{(2)}$ is a decreasing function of $\beta_L$. Also, it is interesting to note that this instability region does
not appear at all if both the phases {\em and} the voltages are initialized to zero!  (See
discussion below for the reason for this behavior.)

\vspace{1em}
Physically, this instability is due to a competition between the self-inductance
of a given loop (say loop $j$), which wishes to have a current with a given sense of
circulation, and the mutual inductance of the two neighboring loops ($j\pm 1$), which wish
to have the current in loop $j$ flow in the opposite sense. 
We have also looked at ladders with $N =6$, $7$, $8$, and $9$.  All show this
instability in the vicinity of $\mu_L =0.5$.  Indeed, we
would expect this competition-induced instability to be independent of ladder size for
the case of nearest-neighbor mutual inductance in that the onset of the instability should
always occur at $\mu_L =0.5$.  

\vspace{1em}
Interestingly, ladders with $N =7$, $8$, and $9$ also exhibit a {\em second}
instability region that the 5-cell ladder does not exhibit.  Fig.~\ref{mutual}(b)
shows the Floquet exponents for $N =7$.  
This second
instability region has an onset at a value of $\mu_{L}^{(3)} > \mu_{L}^{(2)}$ that is
dependent on ladder size. (See cross-hatched region in Fig.~\ref{mutual}(b).) We now turn to an analytic
calculation of the Floquet exponents, which helps us understand the source of these
instabilities.

\vspace{1em}
A reasonable starting approximation is to ignore the coupling between the horizontal and vertical
junctions but {\em otherwise not to ignore the effects of the vertical junctions}.  That is,
we let $\psi =0$ in eq.~\ref{eq:horizphases} and $\phi=0$ in eq.~\ref{eq:vertphases}.  An
analysis like that described in Ref.~\cite{Trees00} applied to eq.~\ref{eq:horizphases} leads to a Mathieu equation
describing the time dependence of the perturbations to the {\em horizontal} Josephson phases.
In such a case, the corresponding Floquet exponents for the horizontal junctions can be 
calculated analytically.  The result is
\begin{equation}
\mbox{Re}\left(\lambda_{m}t_c\right) = -\frac{1}{2\beta_c} \pm \frac{1}{2\beta_c}
\mbox{Re}\sqrt{1 - 4\omega_{m}^{(N)}\left(\frac{\beta_c}{\beta_L}\right)},
\label{eq:mutualexponent}
\end{equation}
where we can think of $\omega^{(N)}_{m}$ as the {\em effective} normal-mode frequencies of
the ladder.  For example, $\omega^{(5)}_{m}=(4\sin^2(m\pi/N) + 2\mu_L\{\cos(2\pi m/N)
-\cos(4\pi m/N)\})/(\mu_L^2 - \mu_L - 1)$, where $0\leq m\leq N-1$.  Note that $\omega^{(N)}_{m}$
is a function of $\mu_L$.

\vspace{1em}
Eq.~\ref{eq:mutualexponent} was used to produce the solid curves in Figs.~\ref{mutual}(a) and (b).
Note that if, for particular values of $\omega_{m}^{(N)}$, $\beta_c$ and $\beta_L$, the 
argument of the square root in Eq.~\ref{eq:mutualexponent} is positive
and larger than one, then at least one of the Floquet exponents will be
positive, signaling unstable phase-locked solutions.  In fact, since $\beta_c > 0$ and
$\beta_L > 0$ such an instability will
occur if $\omega_{m}^{(N)} < 0$!  Plots of 
$\omega^{(N)}_{1}$ versus $\mu_L$ for $N = 5$ and $7$ (not shown here)  demonstrate
that $\omega^{(5)}_{1}>0$ for $0 \leq\mu_L\leq 1$, {\em but} $\omega^{(7)}_1$ is negative
for $\mu_L >0.8$.   
We have checked that $\omega^{(N)}_m > 0$ for $m\neq 1$ and for $N = 5$ and $7$.  Thus
the cause of this instability in the 7-cell ladder for $\mu_L > \mu_{L}^{(3)}$ is the $m=1$ normal mode.
That is, this instability is a geometrical effect, in that it does not occur for $N=5$ for example, and it is triggered by an effective
normal mode frequency of the horizontal junctions becoming negative.

\vspace{1em}
This analytic work, however, does {\em not} point to the horizontal junctions as the cause
of the instability near $\mu_L=0.5$. To appreciate this behavior it is crucial to look
to the vertical junctions.  A procedure similar to that which led to eq.~\ref{eq:mutualexponent}
leads to a set of effective Floquet exponents for the vertical junctions
\begin{equation}
\mbox{Re}\left(\Lambda_{m} t_c\right) = - \frac{1}{2\beta_c} \pm \frac{1}{2\beta_c} \sqrt{1 -
4\beta_c\left[ \alpha\cos\psi_0 + \frac{2\gamma_{m}^{(N)}}{\beta_L}\right]},
\label{eq:vertexponents}
\end{equation}
where the geometrical factor $\gamma_{m}^{(N)}$ is 
also a function of $\mu_L$ and is similar but not identical to $\omega_{m}^{(N)}$.
In this case, the vertical junctions will exhibit an exponentially growing Josephson phase 
if $\gamma_{m}^{(N)} < -(\alpha\beta_L\cos\psi_0)/2.$
Now a plot of $\gamma_{0}^{(5)}$ versus $\mu_L$ (see inset of Fig~\ref{mutual}(a)) shows that the
function abruptly becomes negative at $\mu_L = 0.5$ and asymptotically approaches
zero from the negative side as $\mu_L$ is increased further.  (We have checked that $\gamma_{m}^{(5)} >0$
for $m \neq 0$.  Also, we see similar behavior for the 7-cell ladders.) 
If we assume that $\cos\psi_0 > 0$, then the vertical junctions will be unstable for $\gamma_{m}^{(N)} < 0$.  Based on 
the behavior of $\gamma_{0}^{(5)}$ an instability region will exist for a range
of $\mu_L$ values, $\mu_{L}^{(1)}\leq\mu_L\leq\mu_{L}^{(2)}$ where
$\mu_{L}^{(1)}=0.5$ and $\mu_{L}^{(2)}$ will depend on $\alpha$, $\beta_L$, and
$\cos\psi_0$.  For example, as $\beta_L$ increases we expect that 
$\mu_{L}^{(2)}$ will decrease, {\em i.e.} approach a value of $0.5$.  This
agrees with the behavior of the numerical results for the Floquet exponents.  
Also, the inequality $\gamma_{m}^{(N)} < -(\alpha\beta_L\cos\psi_0)/2$ suggests that the value of $\mu_{L}^{(2)}$
should depend on the value of $\cos\psi_0$.  This is relevant to the
numerical results in Fig.~\ref{mutual}, where we see that the value of $\mu_{L}^{(2)}$ does indeed
on the choice of the starting configuration of
phases and voltages.  In general, then, it is clear that the instability near $\mu_L = 0.5$
originates with the {\em vertical} junctions and would thus be missed by an analysis
that was based solely on the horizontal junctions.  
It is also clear why this instability does not
appear numerically when {\em both} the Josephson phases and the voltages across the junctions are
initialized to zero.  In such a scenario, although the horizontal junctions may be active,
the only possible solution for the vertical junctions is to keep zero voltages and Josephson
phases for all times.  Since we know this instability region is triggered by the vertical
junctions, the vertical junctions have no chance to ``go unstable'' and
thus the instability never appears.

\vspace{1em}
We conclude that mutual inductance between cells of an underdamped ladder array has
the effect of destabilizing synchronization for ranges of values of $\mu_L$, the ratio of
the mutual to self inductance.  These specific ranges of $\mu_L$ that lead to unstable behavior
are geometry dependent.  An analytic calculation of the Floquet exponents based on 
the horizontal junctions agrees with the numerical exponents, based on the full RCSJ
equations, for those values of $\mu_L$ for which stable phase locking occur.  To understand the
cause of all the observed instabilities, however, it is crucial in the analytic work to consider
the behavior of the vertical junctions.  

\vspace{1em}
Although some values of the mutual inductance used in these simulations can not be obtained
in simple ladder arrays, this work suggests that experimentalists may wish to attempt fabrication
of arrays that enhances the mutual over the self-inductance, perhaps making it possible to look
for the rich dynamical behavior predicted here.  Certainly researchers working on the problem
of coherent emission from Josephson junction arrays should be aware this potential for unstable
behavior exists.

\vspace{1em}
The authors wish to thank Barbara Andereck, Tom Dillman, Steve Herbert, Mark Jarrell, and David Stroud for useful
conversations. This research was funded by the Howard Hughes Medical
Institute Undergraduate Biological Sciences Education Program grant
\#71196-529503 to Ohio Wesleyan University.

\begin{figure}
\caption{Ladder array of Josephson junctions with periodic boundary conditions. The horizontal junctions, along 
the rungs of the ladder, are parallel to the $x$~axis, while the vertical junctions are 
parallel to the $y$~axis.    A dc bias current, $I_B$, is injected at each node on 
one side and extracted from the opposite side. The Josephson phase for the horizontal (vertical)
junction in the $j$th plaquette is $\phi_j$ ($\psi_j$). } 
\label{ladder}
\end{figure}

\begin{figure}
\caption{Minimum Floquet exponent for periodic ladders versus the dimensionless, 
nearest-neighbor mutual inductance. (
The results correspond to $i_B =10$, $\beta_c = 10$, and $\beta_L =100$.)
The three different sets of symbols represent three different starting configurations
of voltages across the junctions, while the Josephson phase differences were always
initialized to zero.  The
solid line represents an analytic result (Eq.~\ref{eq:mutualexponent}) based on  
the horizontal junctions. (a) $N = 5$. The analytic result predicts stable phase-locked
solutions for $0\leq\mu_L\leq 1$. The numerical results exhibit an instability, however, for
$\mu_{L}^{(1)}\leq\mu_L\leq\mu_{L}^{(2)}$ where $\mu_{L}^{(1)} =0.5$ and $\mu_{L}^{(2)}$ is
dependent on the starting configuration of phases and voltages, as well as on the value of
$\beta_L$. This instability originates with the vertical junctions.  
INSERT: geometric quantity $\gamma_{0}^{(5)}$ versus $\mu_L$.  (see eq.~\ref{eq:vertexponents})
(b) $N = 7$. In this case, the geometry of the
ladder leads to an instability for $\mu_L > 0.8$ that originates with the horizontal junctions.
This ``large $\mu_L$'' instability is marked in the figure as
a cross-hatched region.  The instability due to the {\em vertical} junctions near $\mu_L = 0.5$ 
still exists but is narrower than for the 5-cell ladder.} 
\label{mutual}
\end{figure}


\end{document}